\documentclass{aa}  
\usepackage{array}
\usepackage{float}    
\usepackage{graphicx}
\usepackage{txfonts}
\usepackage[dvipsnames]{xcolor} 
\begin{document}

   \title{Characterizing the outer disk of eXtended-UV galaxies in the optical domain with deep surveys}

   \author{E. Bernaud\inst{1}\and
          S. Boissier\inst{1}\and
          Junais\inst{2,3}\and
          K. Małek\inst{4}
          \and
          E. Hugot\inst{1}
          \and
          G. Galaz \inst{5}
          }

   \institute{Aix Marseille Univ, CNRS, CNES, LAM, Marseille, France\\
            \email{eloise.bernaud@lam.fr}
            \and Instituto de Astrof\'{i}sica de Canarias, V\'{i}a L\'{a}ctea S/N, E-38205 La Laguna, Spain
            \and Departamento de Astrof\'{i}sica, Universidad de La Laguna, E-38206 La Laguna, Spain
            \and
              National Centre for Nuclear Research, Pasteura 7, 02-093 Warsaw, Poland
            \and 
            Instituto de Astrofísica, Pontificia Universidad Católica de Chile, Vicuña Mackenna 4860, 7820436 Macul, Santiago, Chile\\
             }

   \date{Received , ; accepted , }

 
  \abstract
   {}
   {Extended-UV (XUV) galaxies are galaxies presenting an extended outer ultraviolet disks. Although some Giant Low Surface Brightness (GLSB) galaxies are also XUV, their relation has been seldom studied. This work aims to determine whether a sample of nine XUV galaxies can be classified as GLSBs, by analyzing their photometric properties in deep optical images.}
   {The method presented here uses optical data from the Dark Energy Survey (DES) to construct surface brightness (SB) profiles for each galaxy. A characteristic UV radius is defined to examine the XUV disks. The surface brightness profiles are fitted using simple exponential functions, and the extracted parameters are compared with the literature to identify possible GLSB galaxies. We examined also other diagnostics, including color profiles.} 
   {The analysis of the optical surface brightness profiles of XUV galaxies reveal that they can be classified into three different families. One third meet the GLSB criteria in terms of diffuseness, another third are regular, and the remaining have a stronger decline in the outer disk than in the inner disk, opposite to GLSBs. 
   The color profiles allowed us, in addition to distinguish one of the galaxy, NGC1140, likely resulting from a rare dwarf-dwarf merger.
   }
   {}


\maketitle
%

\section{Introduction}

Low Surface Brightness (LSB) galaxies are a major class of galaxies, as demonstrated by \citet{ONeil_2000} or \citet{2004Minchin} who showed that they could account for about half of the total galaxy population.
Their large number was confirmed in later studies. For example, \citet{2019MNRAS.485..796M} simulations suggest that LSBs are estimated to account for 85$\%$ of the local number density for stellar mass above a threshold of $M_\star > 10^7 \ M_\odot$. The importance of LSB galaxies has increased with this discovery of their number, but also due to the possible new insights into extra-galactic physics. For instance, it was soon realized that various types of LSB galaxies may be dark matter-dominated \citep{Zwaan1995, 1998deblok} or bring constraints on the dark matter nature \citep{kravtsov2024darkmattercontentultradiffuse,2010Lelli}, and they offer an additional constraint for cosmological models in general \citep{2019A&A...622A.103B,Haslbauer_2019}.

LSB galaxies were first defined as galaxies with a disk central surface brightness in the B-band significantly lower than 21.65 mag.arcsec$^{-2}$, the value found by \citet{1970Freeman} for a small sample of regular nearby galaxies. However, later studies \citep{Bothun_1997,1995Mcgaugh} 
demonstrated that this value was biased by selection effects, as predicted by \citet{1976Natur.263..573D}. Several definitions of LSBs can be found in the literature \citep[e.g.,][]{Bothun_1997,1995Mcgaugh,2003Monnier} and they are often described as galaxies with a surface brightness lower than the sky background (around 22.5–23 mag.arcsec$^{-2}$), with variations on the adopted limit and the band.

After early discoveries made with photographic plates, the progresses in observational methods, the use of CCDs and large sky surveys \citep[e.g., SDSS,][]{2009ApJS..182..543A} have increased the number of known LSB galaxies, but also of dedicated studies. It will be still the case in the coming years with the new deep surveys such as the Dark Energy Survey \citep[DES,][]{abbott_2021}, Euclid \citep{euclidcollaboration2024euclidiovervieweuclid} and the Vera C. Rubin Observatory Legacy Survey of Space and Time \citep[LSST,][]{2019ApJ...873..111I}.

As it was quickly realized, a rich variety exist among LSB galaxies. \citet{sprayberry_1995} proposed to distinguish giant LSB (GLSB) based on the measure of the central surface brightness and the scale length of the disk component. They introduced the concept of diffuseness to delimit the GLSBs from more regular LSB galaxies. The most extreme GLSB (that we will use as a reference in our study) is Malin 1 that was unexpectedly found by \citet{1987Bothun} and is still the largest spiral star forming known galaxy. 
This discovery of giant star forming disks called galaxy formation models into question and led to further studies of these galaxies. After Malin 1, other GLSBs galaxies were discovered \citep[see the compilation in][]{sprayberry_1995}, and new GLSBs have been found in the more recent years, with the help of new imaging surveys \citep[e.g.,][]{Hagen_2016,2021MNRAS.503..830S}. 
These galaxies are characterized by their large-scale length, extremely low central surface brightness, but usually also by an important gas mass with $HI$ mass above 10$^{10}$ M$_\odot$. Their formation remains an open question, with multiple hypotheses proposed to explain their origin \citep[e.g.,][]{2016A&A...593A.126B, Hagen_2016, saburova2018}.

Another class of LSB galaxies that has been largely discussed in the recent years \citep{2015VDokkum,Koda1000UDGs,2016Yagi,Mihos_2015,2021Junais} are the Ultra-Diffuse Galaxies (UDGs) a very faint population, defined by their large sizes ($r_{eff}>1.5 kpc$) and very low surface brightness ($\mu_{0,g}>24 mag.arcsec^{-2}$) but are not in the scope of this paper.

While all these definitions were proposed on the basis of the diagnostic made in the optical, after the launch of GALEX \citep{2005GALEX} in 2003, it was quickly discovered that 30$\%$ of regular galaxies present an extended low surface brightness component in the UV and were called XUV galaxies \citep{2005ApJ...627L..29G,Thilker_2007}. These galaxies are characterized by an UV-emitting disk that extends well beyond the optical radius $R_{25}$ defined as the radius where the surface brightness reaches 25 mag.arcsec$^{-2}$ in the B band. This finding revealed ongoing star formation in regions where little or no star formation was previously expected. \citet{Thilker_2007} defined two different types of XUV galaxies. Type 1 XUVs are characterized by structured and bright UV emission located beyond the expected threshold for star formation. Type 2 XUVs have disks that exhibit blue colors (FUV–NIR) across a large, outer region of the disk with low optical surface brightness.

Studies of the UV properties of GLSBs suggested that some GLSBs are, in fact, also XUV galaxies \citep{Boissier_2008,Hagen_2016}.
This finding leads to question the potential link between these two types of galaxies with faint extended disks. Although the UV emission at low surface brightness in GLSB suggested this connection, the definition of XUV galaxies was made by comparing deep UV to less deep optical images (e.g., SDSS). The reverse question : are XUV galaxies systematically GLSB when characterized in the optical is still open. In this work, we propose to tackle this issue and characterize in the optical classical XUV galaxies from \citet{Thilker_2007} for which deeper optical images are now available with DES.

In Section \ref{data} we present the sample selection, the method and the image processing of the DES data we used to compute the surface brightness profiles. Section \ref{results} presents the surface brightness profiles obtained with DES and GALEX data, comparing our method with the literature. We also propose to compute a physically based UV radius and a diagnostic in order to characterize the UV disk of the XUV galaxies. To extend the analysis, we checked the color profiles and spectral energy distribution (SED) fitting results of the sample. In Section \ref{discussion}, we discuss a particular galaxy distinguishing itself from the rest of the sample, and the possible future of this field with new instruments, before concluding in Section \ref{conclusion}.

\section{Sample and data}\label{data}
\subsection{Sample selection}

\begin{figure*}[t]
    \centering
    \begin{minipage}{0.33\textwidth}
           \centering 
           \includegraphics[width=1.\linewidth, height =\linewidth]{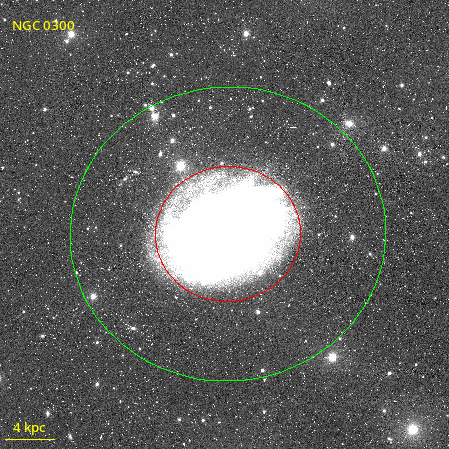}
    \end{minipage}
    \hfill
    \begin{minipage}{0.33\textwidth}
           \centering 
           \includegraphics[width=1.\linewidth, height =\linewidth]{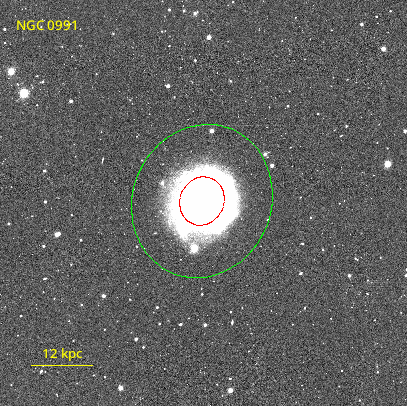}
    \end{minipage}
    \hfill
    \begin{minipage}{0.33\textwidth}
           \centering 
           \includegraphics[width=1.\linewidth, height =\linewidth]{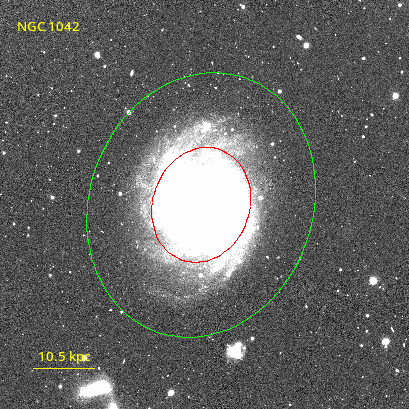}
    \end{minipage}
    \begin{minipage}{0.33\textwidth}
           \centering 
           \includegraphics[width=1\linewidth, height =\linewidth]{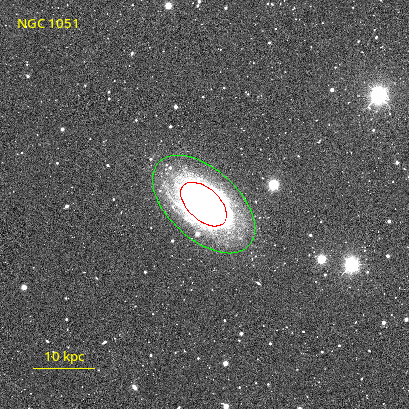}
    \end{minipage}
    \hfill
    \begin{minipage}{0.33\textwidth}
           \centering 
           \includegraphics[width=1.\linewidth, height =\linewidth]{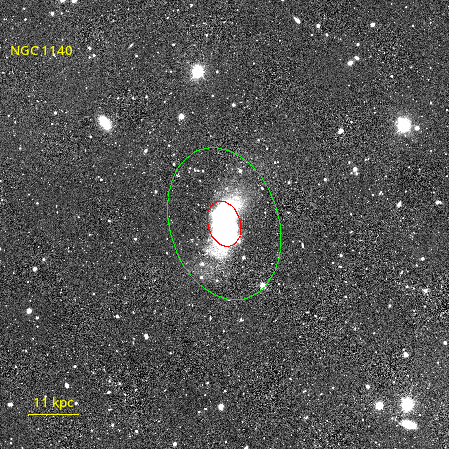}
    \end{minipage}
    \hfill
    \begin{minipage}{0.33\textwidth}
           \centering 
           \includegraphics[width=1.\linewidth, height =\linewidth]{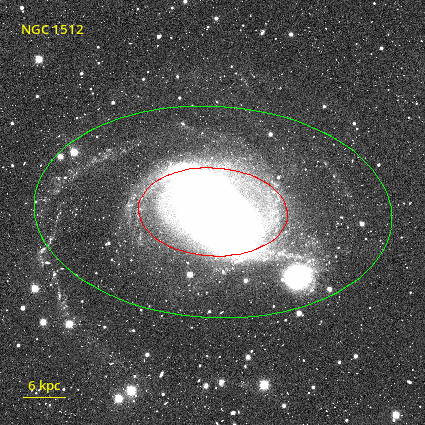}
    \end{minipage}
        \begin{minipage}{0.33\textwidth}
           \centering 
           \includegraphics[width=1.\linewidth, height =\linewidth]{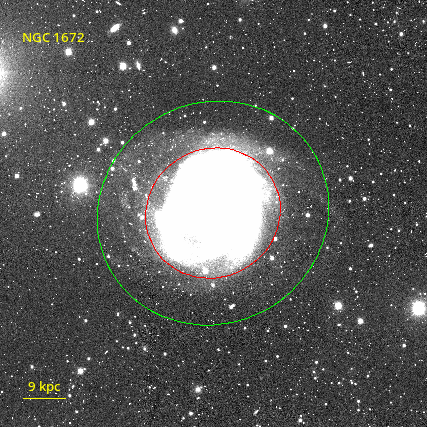}
    \end{minipage}
    \hfill
        \begin{minipage}{0.33\textwidth}
           \centering 
           \includegraphics[width=1.\linewidth, height =\linewidth]{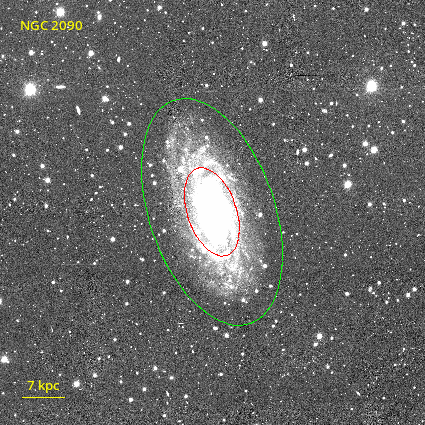}
    \end{minipage}
        \hfill
        \begin{minipage}{0.33\textwidth}
           \centering 
           \includegraphics[width=1.\linewidth, height =\linewidth]{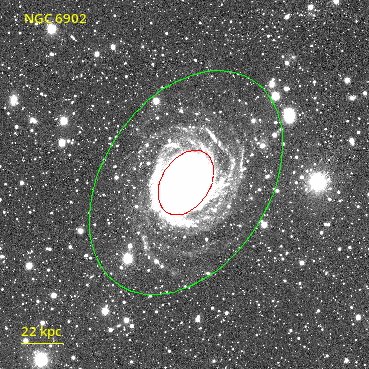}
    \end{minipage}
    \caption{DES images of the sample in \textit{g}-band. The ellipses show the adopted Position Angle and ellipticity at the $R_{25}$ ($R_{UV}$) radius in red (green).}
    \label{fig:DESimages}
\end{figure*}

\begin{table*}[t]
    \centering
    \begin{tabular}{|c|c|c|c|c|c|c|c|c|c|c|}
    \hline
    Object & R.A. & Dec.  & $R_{25}$ & Distance & E(B-V) & Type 1 & Type 2 & $log_{10}(M_{HI})$ & Morphology &scale\\
    Name& degrees & degrees & arcsec & Mpc & & & & $M_{\odot}$ && kpc/arcsec\\
    \hline
    NGC 0300 & 13.722833 & -37.684389 & 609& 2 & 0.0109& Y & N & 9.25 & SA&0.0096\\
    NGC 0991 & 38.886187 & -7.154435 & 48&20.4 &0.0239 &N & Y & 9.15& SABc&0.0989\\
    NGC 1042 & 40.099863 & -8.433544 & 114&18 &0.0245 &Y & N & 9.49&SAB&0.0872\\
    NGC 1051 & 40.260372 & -6.935918 & 54 & 17.1 &0.022 & Y & N & 9.05&SB& 0.0829\\
    NGC 1140 & 43.639922 & -10.027759 & 54 & 18.2 & 0.0318& Y & Y & 9.34&Ibm&0.0882\\
    NGC 1512 & 60.976167 & -43.348861 & 210 & 10.4 &0.0091 & Y & N & 9.46&SB&0.0504\\
    NGC 1672 & 71.427093 & -59.247267 & 192 & 15.1 & 0.02& Y & N & 10.0&SB&0.0732\\
    NGC 2090 & 86.757875 & -34.250611 & 129 & 11.3 &0.0344 & N & Y & 9.35&SA&0.0547\\
    NGC 6902 & 306.11725 & -43.653528 & 99 & 37.6 &0.0346 & Y & N & 10.2&SA&0.182\\
    \hline
    \end{tabular}
    \hspace{1cm}
    \caption{Properties of the sample. Right Ascension (R.A.), Declination (Dec.) and the morphology come from NED, the $R_{25}$, the Distance, the XUV types and the $log_{10}(M_{HI})$ are from \cite{Thilker_2007}. E(B-V) is taken from \protect\url{https://irsa.ipac.caltech.edu/applications/DUST/}}
    \label{tabproperties}
\end{table*}

In order to characterize in the optical the regions detected in UV in XUV galaxies, we cross-correlated the \citet{Thilker_2007} catalog of 54 classical XUV galaxies with the publicly available deep optical images from  Dark Energy Survey \citep[DES][]{abbott_2021}. We analyze in this paper the nine XUV galaxies from \citet{Thilker_2007}  with available deep DES optical ($g$, $r$, $i$, $z$, $Y$) from DR2 and GALEX (FUV and NUV) images. The main properties of the sample are reported in Table \ref{tabproperties}. This small sample includes both type 1 and type 2 XUV galaxies and covers broadly the RC3 T-type of the parent sample of \citet{Thilker_2007}. The morphological type is given in Table \ref{tabproperties}. Some GLSBs have been found in previously misclassified early-type galaxies \citep{Hagen_2016}. We take all the XUV galaxies from \cite{Thilker_2007} independently of their morphological type.

\subsection{Image processing}\label{method}

We produced cutout of DES and GALEX images with a width of 800 arcsec (except for NGC300 for which we used 4000 arcsec by coadding four tiles of DES) in order to ensure in each case to probe the extended disk and keep background sky in the close surrounding of the galaxy. For the GALEX images, we took the archival images available in the GR6/7 Data Release (\url{https://galex.stsci.edu/gr6/}), and we stacked the multiple exposures within the field of the cutout using the \textit{reproject} package \citep{thomas_robitaille_2023_7584411}. We projected the GALEX images with a pixel scale of 1.5 arcsec pixels to the DES pixel scale of 0.263 arcsec using again the \textit{reproject} tool.

Some galaxies (in particular NGC1042) have bright neighbors. This causes fluctuations and flux contamination in the images and increases the uncertainties in the faint and extended parts. The first step to limit this effect is to mask all sources in the g+r images including the galaxy itself using \textit{segmentation} from \textit{photutils} \citep{larrybradley2024}. We then fit the masked image with a second order polynomial as in \citet{Merritt_2016}. This sky gradient is subtracted from the images.

In order to estimate the sky characteristics, we created 24 boxes (of size varying from 70 arcsec to 437 arcsec depending on the size of the cutout) around each galaxy. In each of them, we computed the mean ($\mu$) in each box in the masked images (using the g+r mask created before). The boxes with a mean greater than three sigma from the average of the means ($<\mu>$) were excluded to avoid boxes with badly masked bright stars or artifacts. 

The masked pixels were replaced by the mean sky value ($<\mu>$) calculated above.
We then applied a 13x13 pixels binning.
This value was chosen by comparing the binning applied by \citet{Pohlen_2006} to NGC1042 with the SDSS data deferred to the DES pixel size. 

Field stars overlapping with the galaxy were  replaced by the mean value measured in a circular annulus surrounding them using the function \textit{SkyCircularAnnulus} of \textit{photutils}.

Some faint artifacts or stars made visible by the binning were masked again in the binned image, that we finally used to compute the sky properties in the same boxes as above: the mean of the number of pixels (NPix), mean of the means ($<\mu>$), the mean of the standard deviation in each box ($e_{p}$) and the standard deviation of the means ($e_{s}$). These constants allow us to compute surface brightness uncertainties (used in section \ref{sbprofiles}) following Equation \ref{equationerror} from \citet{Gil_de_Paz_2005}:

\begin{equation}
    \sigma_{SB} = \sqrt{E_{1}+E_{2}} 
\label{equationerror}
\end{equation}

with 
\begin{equation*}
        E_{1} = \frac{e_{p}^2}{Ndata} \hspace{0.2cm} and \hspace{0.2cm} E_{2} = max(e_{s}^2 - \frac{e_{p}^2}{Npix},0) 
\end{equation*}

Ndata represents the number of data points.
This way of computing errors is as conservative as possible, taking into account both the normal distribution on the pixel scale measured by $e_p$ and the eventual variations on large scales, measured by $e_s$, while errors in the literature often refer only to the normal distribution in carefully chosen regions.

\section{Analysis and results}\label{results}
\subsection{Surface brightness profiles}\label{sbprofiles}


   \begin{figure*}[t]
   \centering
      \includegraphics[width=1\textwidth,height = 1\textwidth]{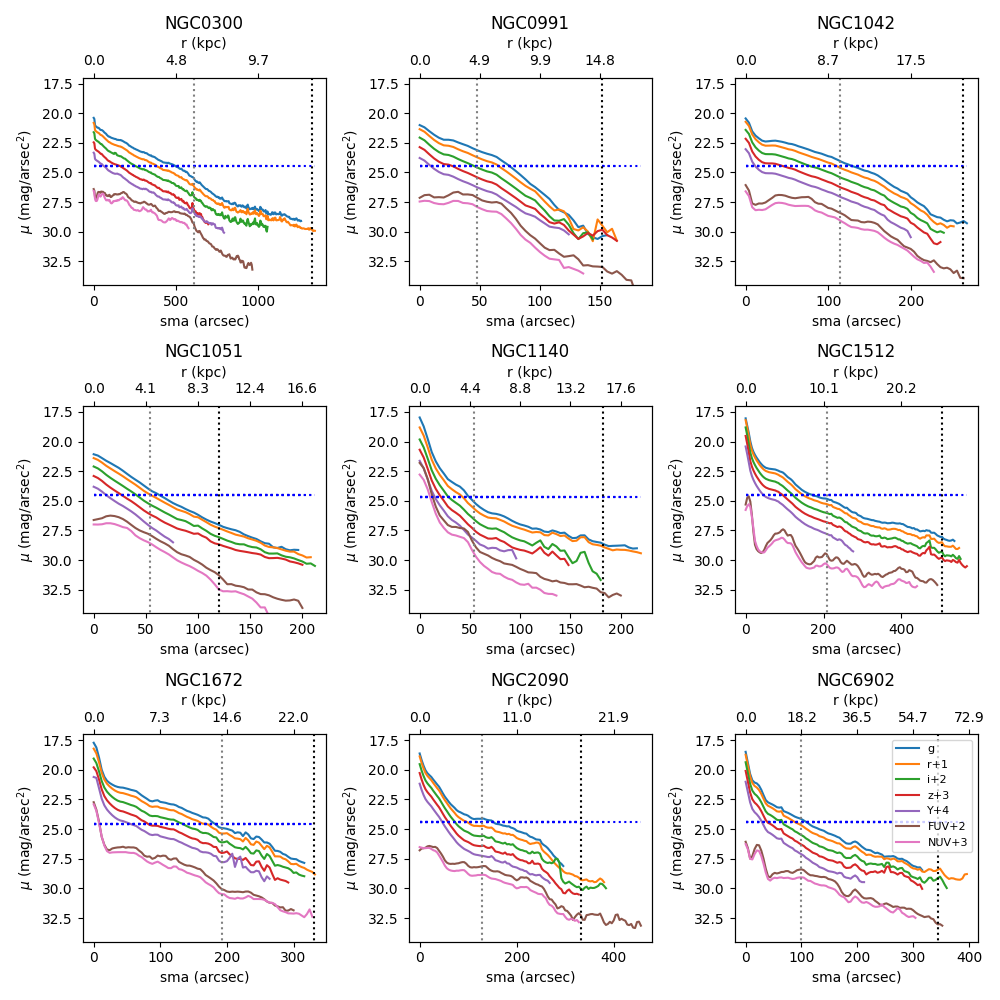}
      \caption{The observed Surface Brightness Profiles for each band of the sample as a function of the semi major axis of the ellipses. The gray dashed line is the $R_{25}$ and the black dashed line is the $R_{UV}$ defined in the Section \ref{RUV}. For clarity, an offset have been applied on the profiles of +n magnitudes/arcsec$^2$ with a depending on the bands indicated in the bottom right panel. The blue dotted line represents 25 mag.arcsec$^{-2}$ in the B-band adapted to the \textit{g}-band using \cite{1995PASP..107..945F}.}
      \label{Figsallbands}
   \end{figure*}
   \begin{figure}[t]
   \centering
      \includegraphics[width=0.5\textwidth]{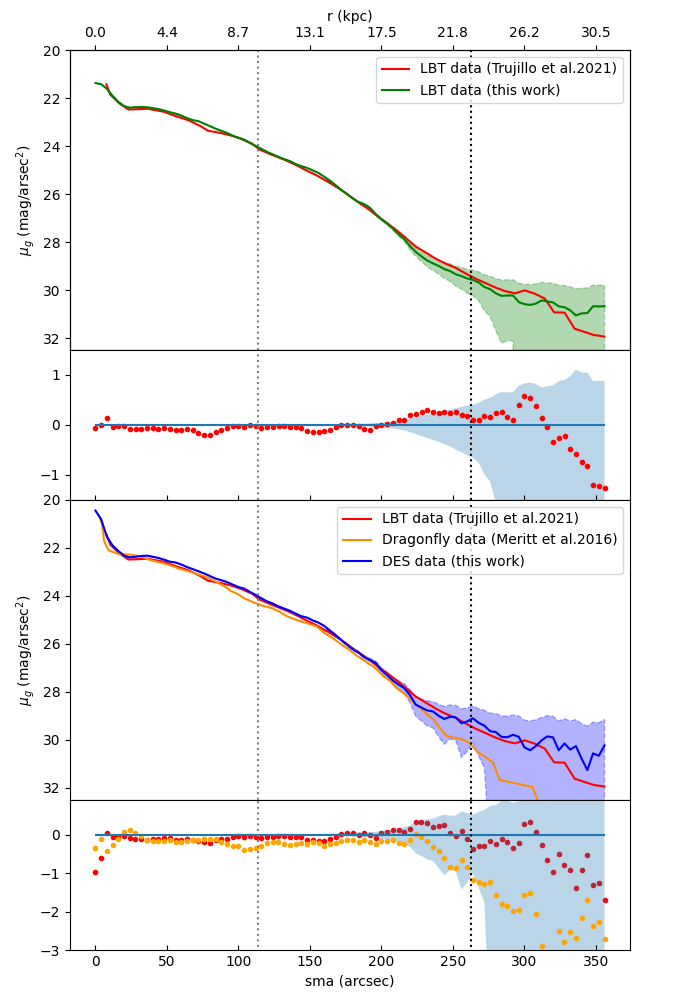}
      \caption{Surface brightness profile in \textit{g}-band for NGC1042 as a function of the semi-major axis (sma) of ellipses. \textbf{Top:} Comparison of the method with the profile from \cite{Trujillo_2021} with LBT data (in red) and our method applied to the same data (in green) and the difference between them under the main panel. \textbf{Bottom:} Comparison of the data with the profile from \cite{Trujillo_2021} with LBT data (in red), the profile from \citet{Merritt_2016} with Dragonfly data and our method applied to DES data (in blue). The difference between the other profiles and our are shown under the main panel. The gray dashed line is the $R_{25}$ of the galaxy when the black dashed line is the $R_{UV}$.The shaded area shows the uncertainties for LBT data with the method used here in the top panel and DES data with the method used here in the bottom panel.}
              
      \label{FigSBg}
   \end{figure}

In order to obtain surface brightness profiles at the same location in all the bands for each galaxy, the Position Angle (PA) and ellipticity are fixed following the idea of \citet{Pohlen_2006} to determinate these parameters in a deep image, adequate for the outer parts of the disks. 
It has the following advantages : in the outer regions with a low signal, it avoids the PA to change arbitrarily, and it is a method consistent with many surface brightness profiles from the literature \citep{Trujillo_2021,Merritt_2016,Pohlen_2006}. We also performed in which we let the PA and ellipticity vary freely for NGC1512 for which the geometry varies the most with radius. The results of this test were consistent with our method and the conclusions of this work remained unchanged.

We thus first generate a set of ellipses in the $g$ band with \textit{EllipseGeometry} and \textit{Ellipse} from the \textit{photutils} package with the center fixed, the position angle (PA) and the ellipticity (\textit{e}) set as free parameters. For these free parameters, a guess is provided, but we verified that it has no impact on the final results. 
We then select the ellipse that reaches a surface brightness one sigma above the sky mean, following \citet{Pohlen_2006}. In this way, we keep the ellipses as close as possible to the shape of the external disk of the galaxy. 

The resulting PA and $e$ are given in Table \ref{tabcalculs}. Then a new set of ellipses with a spacing of 4 arcsec was defined with these parameters fixed. This set of ellipses is used to measure the surface brightness profiles in each band with the function \textit{fit\_isophote} from \textit{Photutils} and are shown in Fig. \ref{Figsallbands}. 
The profiles are cut whenever the measured profile reaches the sky value or the errors are too large (above one magnitude). The complete profiles with their uncertainties are provided in an electronic form \footnote{\url{https://docs.google.com/spreadsheets/d/1qulS5S\_tyI3MH71Am42e4PKFF7xQyC2KpdqKKLBYzBw/edit?usp=sharing}}. We compared the GALEX profiles obtained with other profiles available in the literature. We found that our profiles are similar to those published by \cite{2018ApJS..234...18B} but generally extend to larger radii than those analyzed in \cite{Gil_de_Paz_2007}.

\subsection{Robustness of the method}

Among our nine galaxies, NGC1042 was observed in the $g$ band with the Large Binocular Telescope \citep[LBT,][]{Trujillo_2021} and Dragonfly \citep{Merritt_2016} to study its outer parts and stellar halo. This gave us the opportunity to perform several tests to ensure that the DES images are deep enough (and our method robust enough) to study the region of the galaxy responsible for the extended UV emission.

First, we computed the surface brightness with the LBT data using the method described in Sect. \ref{method} and \ref{sbprofiles}.
We obtained a very similar profile to the one from \citet{Trujillo_2021}, calculated with the same data but with a different approach (see the top panel of Fig. \ref{FigSBg}).
Our error bars are larger than theirs, which can be attributed to the more conservative approach adopted in calculating uncertainties that include a large-scale fluctuations component (see Equation \ref{equationerror}). 

\begin{table*}[t]
    \centering
    \begin{tabular}{|c|c|c|c|c|c|c|c|c|c|c|}
    \hline
    Object & P.A  & \textit{e} &$R_{UV}$ & $\mu_{0,inner}$   & $\mu_{0,outer}$ & $R_{s,inner}$ & $R_{s,outer}$ & $L_{R25}$ &  $L_{RUV}$ & $d_i$(outer)\\
    Name& degrees & & arcsec & mag & mag & arcsec & arcsec& mag abs & mag abs&  \\
    \hline
    NGC 0300 &4.93 & 0.069 & 1325& 22.46 & 24.81 & 159.51 & 242.27  & -18.17 & -18.26 &25.89 \\

    NGC 0991 & 68.0 & 0.11 & 152 & 21.97& 19.38 & 28.06& 13.5 & -18.67 & -19.29 &19.2\\

    NGC 1042 & 71.0 & 0.17 & 263 & 22.22 & 19.78 & 52.06 & 26.87 & -19.81 & -20.11 &20.86\\

    NGC 1051 & 137.5 & 0.43 & 120 & 22.33 & 22.78 & 18.76& 24.16 & -18.32 & -18.65 &23.51\\

    NGC 1140 & 106.69 & 0.3 & 182 & 20.03 & 25.07 & 9.41 & 45.62  & -19 & -19.12& 27.31\\

    NGC 1512 & 176.74 & 0.41 & 504 & 21.47 & 24.22 & 46.02 & 112.14 & -19.84 & -20.05 & 27.21\\

    NGC 1672 & 29.34 & 0.06 & 330 & 20.64 & 19.75& 44.41 & 41.15& -20.89 & -20.98 &21.37\\

    NGC 2090 & 109.0 & 0.47 & 332 & 21.87 & 22.19  & 27.01& 42.1& -19.53 & -19.85 & 23.23\\

    NGC 6902 & 56.86 & 0.32 & 344 & 21.63 & 23.95  & 25.04& 61.14& -21.33 & -21.72 &28.41\\

    \hline
    \end{tabular}
    \hspace{1cm}
    \caption{Properties of the sample computed for this work. The Position Angle and the ellipticity are chosen as described in Section \ref{method}, the $R_{UV}$ from Section \ref{RUV}. The PA is defined in this work from W to N.The central surface brightness and the scale length are computed as in Section \ref{results}. Typical uncertainties on $\mu_{0,g}$ are a few tenth of the magnitude while the uncertainties for the $r_s$ are lower to 10$\%$. The luminosity is calculated from the $g$-band profile within $R_{25}$ and $R_{UV}$. $d_i$ is the diffuseness calculated from the equation \ref{EqDiffusness}, following \citet{sprayberry_1995}}.
    \label{tabcalculs}
\end{table*}

Second, we compared the surface brightness profile obtained with the DES data and our method with the \citet{Trujillo_2021} LBT surface brightness profile (bottom panel of Fig. \ref{FigSBg}). This step allowed us to verify that our method allows us to measure a surface brightness profile up to the largest radii measured by \citet{Trujillo_2021}, here again with large uncertainties at the largest radii. Small differences between the data are visible just before the UV radius (defined in Section \ref{RUV}), and do not affect our results.
We show one sigma upper limits in the profiles whenever the measured flux minus the uncertainty is lower than zero. The uncertainties are slightly larger than those obtained with the LBT data, which is directly related to the relative noise of the LBT and DES images. Importantly, the uncertainties remain small in the "XUV region" that we want to analyze, and are under one $\text{mag.arcsec}^{-2}$ at its outer edge (see section \ref{XUVrelation} for a definition of our maximum radius of analysis). 

We also compared our profiles with the one obtained by \citet{Merritt_2016} with Dragonfly for this galaxy (lower panel of the Fig. \ref{FigSBg}).
We note that the Dragonfly profile drops to a surface brightness lower than the DES or LBT profiles at the largest radii, but our conservative uncertainties encompass it.

As a further test, we applied a coarser binning in DES images of 65x65 pixels to obtain a similar pixel scale as the dragonfly value of 2.85 arcsec/pixel with the binning they used, running a first set of ellipses on $g+r$ images (instead of $g$ in our case) and using those ellipses to compute the $g$ band profile.
The profile obtained in this procedure is consistent with the one obtained above, without improving it.  
As shown in Fig \ref{Figsallbands}, some of the profiles flatten in the very outer parts, particularly for NGC991, NGC1042 (see also Fig. \ref{FigSBg}), NGC1140, NGC2090. This flattening begins when the flux is close to the sky value and the uncertainties are high).

We thus consider that our data and methods are robust in comparison to other data sources and methodology to produce a surface brightness profile in the UV emitting regions, and we adopt it for the rest of this study. 

All data in the following sections have been corrected for inclination by multiplying the fluxes by $$(1-e) = \frac{b}{a}.$$ The data hereafter are also corrected for Galactic extinction $A_{\lambda}$. For each band, the extinction law $<\frac{A_{\lambda}}{A_{V}}>$ is calculated using \cite{1989ApJ...345..245C}. Then we have subtracted from the profiles the attenuation $$ A_{\lambda} = \frac{A_{\lambda}}{A_{V}} \times R_{V} \times E(B-V)  $$ $R_{v}$  is set to a value of 3.1, and the reddening E(B-V) is given in Table \ref{tabproperties}. 

According to our profiles, the $R_{25}$ seems to be further out than the one adopted in \citet{Thilker_2007}, as can be seen from Figure 2. In order to check how much our results depends on the determination of $R_{25}$, we calculated our own value based on the \textit{g}-band profile, converted to B-band using \citet{1995PASP..107..945F}. We found small changes (between six and thirteen arcsec) in most cases, except for two galaxies (117 arcsec for NGC0300, 20\% smaller, and 24 arcsec for NGC 0991, 50\% larger). We verified that our results and figures are qualitatively unchanged if this value is adopted. For consistency with previous works, we show our results adopting the value of $R_{25}$ from the literature for the rest of the analysis. 

\subsection{Characterization of the XUV galaxies}\label{XUV}

\subsubsection{Determination of $R_{UV}$}\label{RUV}

The size of galaxies in the UV is frequently defined as the last galaxy radius with useful UV data \citep{2007ApJS..173..185G,Boissier_2008}. This definition is not satisfactory as it depends on the data, especially with our goal of characterizing the UV extended region. 
We thus propose to adopt a definition that is consistent across all galaxies and physically motivated, by basing it on the star formation rate surface density. 

To do so, we converted the inclination and Galactic extinction corrected surface brightness profile in the FUV band of GALEX in a flux $F_{\nu}$ in $ergs$ $s^{-1} Hz^{-1} cm^{-2}$, and then into a star formation rate surface density following
Equation \ref{equationSSFR}:

\begin{equation}
    \Sigma_{SFR} = F_{\nu} \times C  \times  5.08  \times10^{57}  \ \ \  in \ \ M_{\odot} Gyr^{-1} pc^{-2}
\label{equationSSFR}
\end{equation}

\begin{figure}[t]
\centering
  \includegraphics[width=0.5\textwidth,clip]{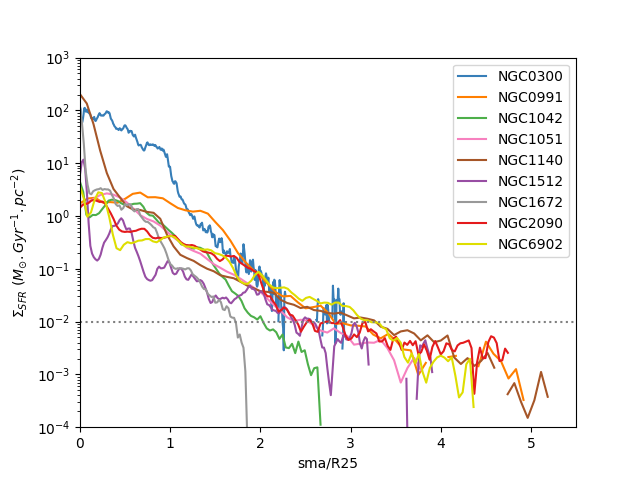}
  \caption{Star Formation Rate Surface Density ($\Sigma_{SFR}$) as a function of the semi major axis normalized to the $R_{25}$. The gray dashed line is the value corresponding to the $R_{UV}$.
  }
  \label{FigRUV}
\end{figure}

The $C$ constant ($0.97 \times 10^{-28}$) represents the UV-SFR calibration, for which we adopt the value of \citet{boissier2013} using the IMF of \citet{kroupa2001}. The results are shown in Fig. \ref{FigRUV}. 
We then adopted for $R_{UV}$ the radius where the SFR surface density is equal to
$\Sigma_{SFR}=10^{-2} \ M_{\odot} Gyr^{-1} pc^{-2}$. This value is well reached in all the GALEX profiles and it corresponds to what is typically found in the very outer regions of nearby galaxies studied in \citet{Bigiel_2008}, and to the value far out in the disk of Malin 1 \citep{Junais_2024}. This radius is thus well consistent with the largest star-forming regions observed in galaxies in general. The $R_{UV}$ obtained for each galaxy in our sample is given in Table \ref{tabcalculs}. Fig. \ref{FigRUV} shows that $R_{UV}$ in our XUV galaxies is, of course, much larger than $R_{25}$, ranging from about 2 to 3.5 $R_{25}$.



\subsubsection{Optical characterization of the extended UV region}\label{XUVrelation}

In this section, we aim to characterize in the $g$ band the regions where UV emission is found at large radius in XUV galaxies. Large diffuse disks have been classically studied in the optical. \citet{sprayberry_1995} proposed a diagnostic plot showing the central surface brightness vs the scale-length of the disk component of such galaxies.

In order to add the XUV selected disks in this (optical) diagnostic plot we applied the following procedure.
The profiles are fitted with simple exponential functions in the $g$-band in the inner region and outer regions of our galaxies.

\begin{figure}[t]
\centering
  \includegraphics[width=0.5\textwidth,clip]{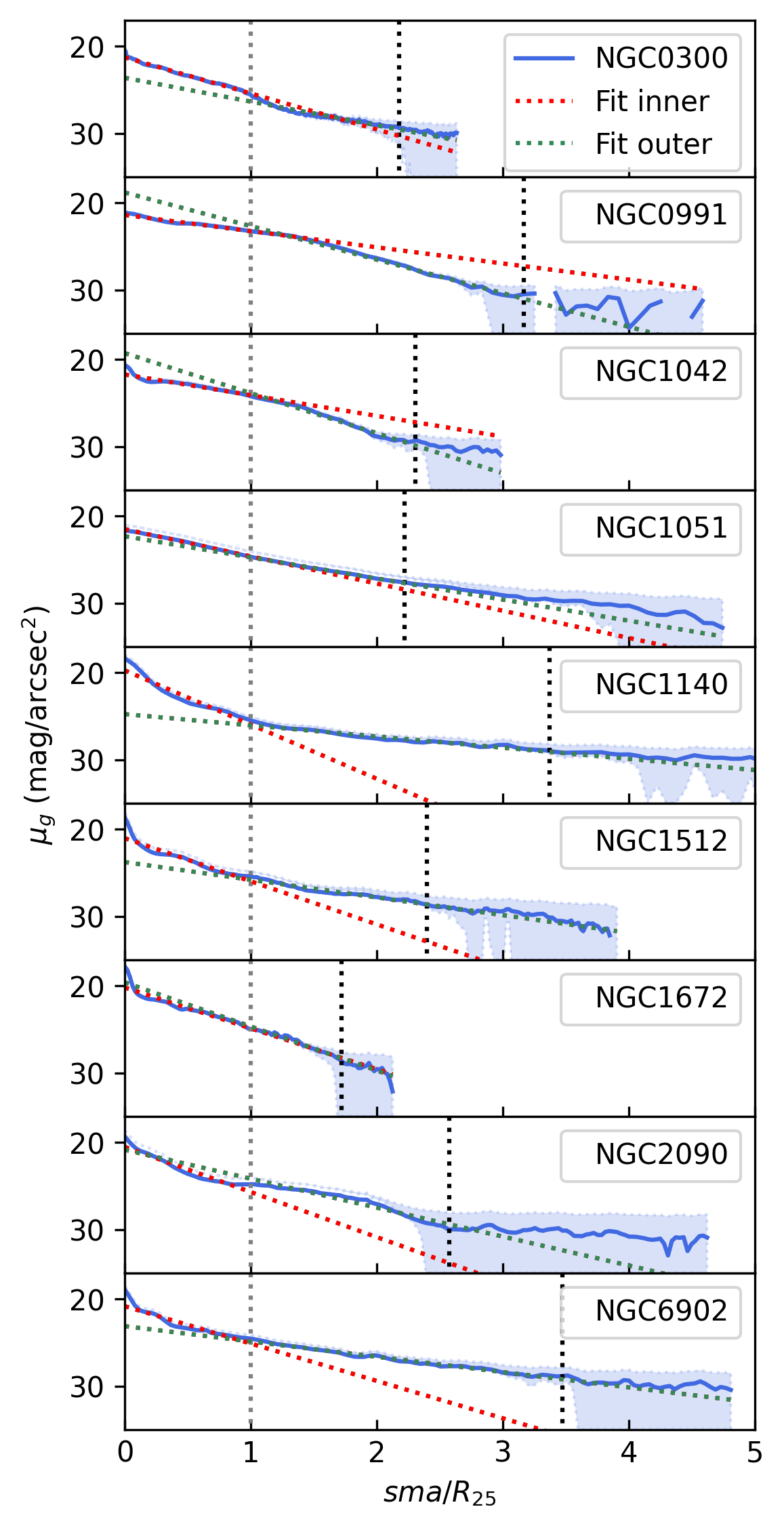}
  \caption{Surface brightness profiles in g-band of the sample as a function of the radius  normalized by $R_{25}$. The gray dashed line is the $R_{25}$ and the black one is the $R_{UV}$. The green dot line is the fit in the outer region and the red dot line is the fit in the inner region.
  }
  \label{Figsfits}
\end{figure}

The inner region is defined as the area within the radius $R_{25}$ of the galaxy, while the outer region corresponds to radii ranging from $R_{25}$ to $R_{UV}$, the latest being defined in Sect. \ref{RUV}. For each galaxy, we applied an exponential fit to the radial profile in each region and show the results in Fig. \ref{Figsfits}. For each region, we thus obtained the central surface brightness and the scale length ($\mu_{0,g}$ and $R_{s}$) that are given in table \ref{tabcalculs}. 
This procedure does not necessarily provide a perfect fit (see Fig. \ref{Figsfits}), but is extremely simple and easy to reproduce.
A bulge-disk decomposition could be more effective. We tried this approach by fitting two components exponential and De Vaucouleur profiles. We obtained very similar scale lengths and central surface brightness of the disk component for 6 galaxies. However, the remaining three galaxies (NGC6902, NGC2090, NGC1140) were dominated by an early-type profile at all radii inside $R_{25}$ showing that this approach is difficult to apply systematically. We also tried to determine these parameters after excluding the most inner points that may be affected by a bulge, and found very similar values.

\begin{figure}[t]
\centering
  \includegraphics[width=0.5\textwidth,clip]{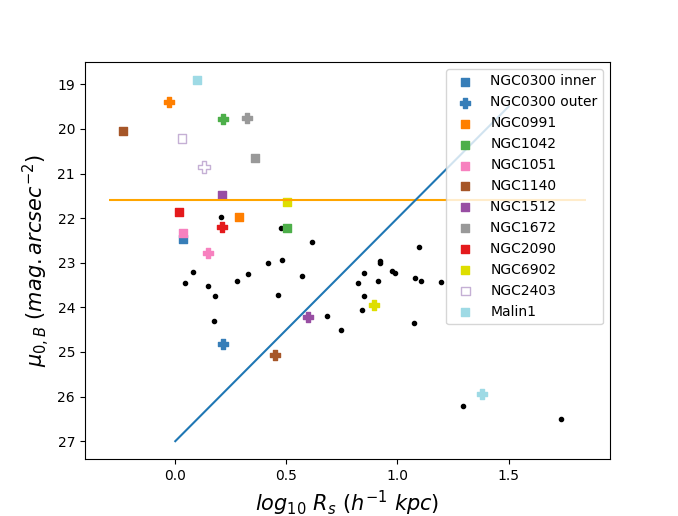}\caption{Classification of the sample adapted from \citet{sprayberry_1995, Hagen_2016}. The squares symbols represent the value of the disk fit for the inner part of the galaxies and the cross symbols for the extended UV region. The blue line distinguishes the Low Surface Brightness galaxies (LSB) from the Giant LSB (GLSB) galaxies. The orange line, the \cite{1970Freeman} value (21.65 $\text{mag.arcsec}^{-2}$), separates the high surface brightness (HSB) galaxies from the LSB galaxies. In order to compare with the data of \citet{sprayberry_1995,Hagen_2016,1994AJ....107..530M} (black dots) calculated with $H_{0} = 100 \ h \ km \ s^{-1} Mpc^{-1}$ we corrected the Scalelength with $h^{-1}$ } 
  \label{Figsprayberry}
\end{figure}

In order to compare to \citet{sprayberry_1995}, the profiles were converted from the g-band to the B-band, following \citet{1995PASP..107..945F}.
While Malin-1 was present among the galaxies from \citet{sprayberry_1995}, we decided to fit its profile with exactly the same procedure.
We used the I-band profile from \citet{2007AJ....133.1085B} converted in g-Band using the conversion of \citet{1995PASP..107..945F}, and determined $R_{UV}$ from the FUV profile of \citet{2016A&A...593A.126B}. 
We obtained a value for the outer region of Malin 1 close to the value found by \citet{sprayberry_1995} with a different decomposition of the profile (the most extreme point in the Figure \ref{Figsprayberry}). Our outer region scale-length is slightly smaller, but still in the same area of the diagnostic plot.

Figure \ref{Figsprayberry} shows that our sample is located in different part of the Sprayberry diagnostic plot and three families can be distinguished as described below.

The first family is defined as the \emph{Malin 1-like} galaxies. It is composed of NGC6902, NGC1512 and NGC1140, for which the outer region falls within the GLSB region, while the inner region is in the HSB area. Thus, one third of our XUV galaxies could be considered GLSBs.
However, GLSBs are sometimes defined with additional criteria with respect to the Sprayberry plot. For instance, \citet{2023MNRAS.520L..85S} add a criteria on their size that is fulfilled only by NGC6902 among the three galaxies. \citet{2023MNRAS.523.3991Z} used instead a criteria on the HI mass (above $10^{10} M_\odot$) and size (larger than 50 kpc). Only NGC6902 again is sufficiently massive to be a potential GLSB according to this definition. The Malin-1 like family described in this paper refers only to the position in the Sprayberry plot, and its diffuseness (see below).

The second family, the \textit{regular} galaxies, is composed of NGC1051, NGC2090 and NGC0300. The outer central surface brightness ($\mu_{0,outer}$) is fainter than the inner region one, as for the Malin1-like, but the outer disc is not in the giant LSB area, with a smaller scale-length. The inner and outer values of $\mu_0$ and $R_s$ are similar, the profiles not changing much at $R_{25}$. We also notice that the inner regions for the three galaxies are on the LSB side of the figure \citep[$\mu_{0,inner}$ below the][value]{1970Freeman}.  

The last family is made of NGC1042, NGC0991, NGC1672 for which both the inner and outer regions are in the hight surface brightness (HSB) area and $\mu_0$ is fainter in the inner region than in the outer region. We want to stress that the fact that the outer disks fall in the HSB area does not mean that the optical disk is HSB. Actually, by construction, the disks are themselves below 25 $\text{mag.arcsec}^{-2}$. A high central surface brightness of $\mu_{0,outer}$ is rather indicative of a down-bending profile \citep{Pohlen_2006}. In our case, this change being found between the inner disk and the XUV region.
We call this family the \emph{Malin-1-opposite}. 

Even if our sample is too small to derive definitive conclusions, we note that there is no systematic correlation between the families described above and the XUV type (1 or 2), or the shape of the profile (exponential, up-bending, down-bending) defined by \citet{Pohlen_2006}, although \textit{Malin-1 like} galaxies are not down-bending and the \textit{Malin-1 opposites} are not up-bending.

To better characterize optical disks of our sample of XUV galaxies, we used a 'diffuseness index' defined by \citet{sprayberry_1995}. GLSBs are separated from regular LSB galaxies at a diffuseness index of 27 and designated as 

\begin{equation}\label{EqDiffusness}
    d_i= \mu_{0,B} + 5*log(R_{s})
\end{equation}

We computed the diffuseness of the outer disk of each galaxy (see Table \ref{tabcalculs}). We found that diffuseness correlates (with some dispersion) both with the scale-length and the central surface brightness of the outer disc, what is expected from its definition.

We could not establish any other link with previously measured values such as $\mu_{0}$ or $R_{s}$ of the inner fit. The diffuseness does not seem to correlate with other global values (mass of gas, integrated $g$-band luminosity from the profile, either integrated within $R_{25}$ or $R_{UV}$, given in Table \ref{tabcalculs}). On average, the Type 1, 2 and 1+2 XUV galaxies have a mean diffuseness of 26, 23 and 24 respectively. This indicates that type 1 XUV galaxies may be more diffuse in the optical regime than type 2 ones, but the size of our sample is too small to derive strong conclusions on this point.

We also explored if the difference between the fit values in the inner vs outer part could be related to other properties. For truncated galaxies, we expect $\mu_{0,inner}$ > $\mu_{0,outer}$ and $R_{s,outer}$< $R_{s,inner}$, while it is the opposite for anti-truncated galaxies. Note that this definition is not always in line with the description of the literature of truncated (down-bending) or anti-truncated (up-bending) galaxies. Surface brightness mostly because in our case we are focused on the difference between the XUV area (outer region, $R_{25} < R < R_{UV}$) and the disks inside $R_{25}$ (inner region), while the truncation nature of the profiles in the literature is usually considered on the basis of the full profile, without focusing on a particular radial range.

Again, we did not find any strong correlation that would relate the difference between inner and outer disks to other properties of the galaxies. However, a correlation is expected by definition, once it is noted that the two exponential fit should be continuous at $R_{25}$, i.e.,

$$\mu_{inner}(R_{25}) = \mu_{outer}(R_{25}) = 25 \  \text{mag.arcsec}^{-2},$$

what leads to the relation:
\begin{equation}
    \mu_{0,outer} -  \mu_{0,inner} = 2.5 \times log(10) \ DIS 
    \label{equationmurs}
\end{equation} 

Where DIS stands for the difference of inverse scalelength, defined as $$ DIS = (\frac{R_{25}}{R_{s,inner}}-\frac{R_{25}}{R_{s,outer}})$$

\begin{figure}[t]
  \includegraphics[width=0.55\textwidth,clip]{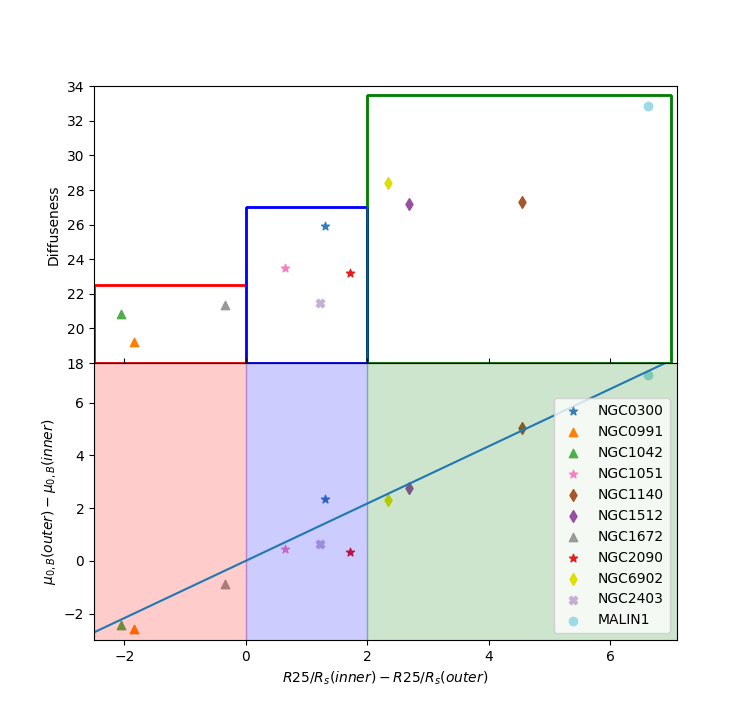}
  \caption{\textit{Top :} The diffuseness as a function of DIS. \textit{bottom :}The difference between the central surface brightness in the outer and inner region as a function of DIS. The blue line represent the Equation \ref{equationmurs}. Each dots correspond to a XUV family triangle: Malin-1 opposite; diamond: Malin-1 like; stars: regular}
  \label{Figsmu0rs}
\end{figure}

Although the fits are very simple (pure exponential, without decomposition), not perfect and not forced to reach exactly 25 at $R_{25}$, the values we obtain are consistent with Equation \ref{equationmurs}, as shown in Fig. \ref{Figsmu0rs}. The three different families are actually simply separated by their DIS : The \textit{Malin1-opposite} are found for negative values, the \textit{regular} galaxies are found at $0<DIS<2$. Malin-1 like (and Malin 1) are found above two. Since our galaxies follow Equation \ref{equationmurs}, the three families are also distinguished by their value of $\mu_{0,outer} -  \mu_{0,inner}$.  
Furthermore, we found a strong correlation between the DIS and the diffuseness of the outer part: the \emph{Malin1-opposite} are found under a diffuseness index of 21, the \emph{regular galaxies} are located between 21 and 27, and the \emph{malin1-like} are above 27. The families we defined can thus be broadly distinguished using only one parameter (e.g., DIS or diffuseness).

Note, however, that this conclusion applies only to the structural properties obtained with the two exponential fits; we will see in the following that NGC1140 distinguishes itself when the color gradient, stellar age radial variation, or morphology are considered, as discussed in Sect. \ref{NGC1140}.

\subsection{Color profile and SED-fitting}\label{sedcolor}

\begin{figure}[t]
\centering
  \includegraphics[width=0.5\textwidth]{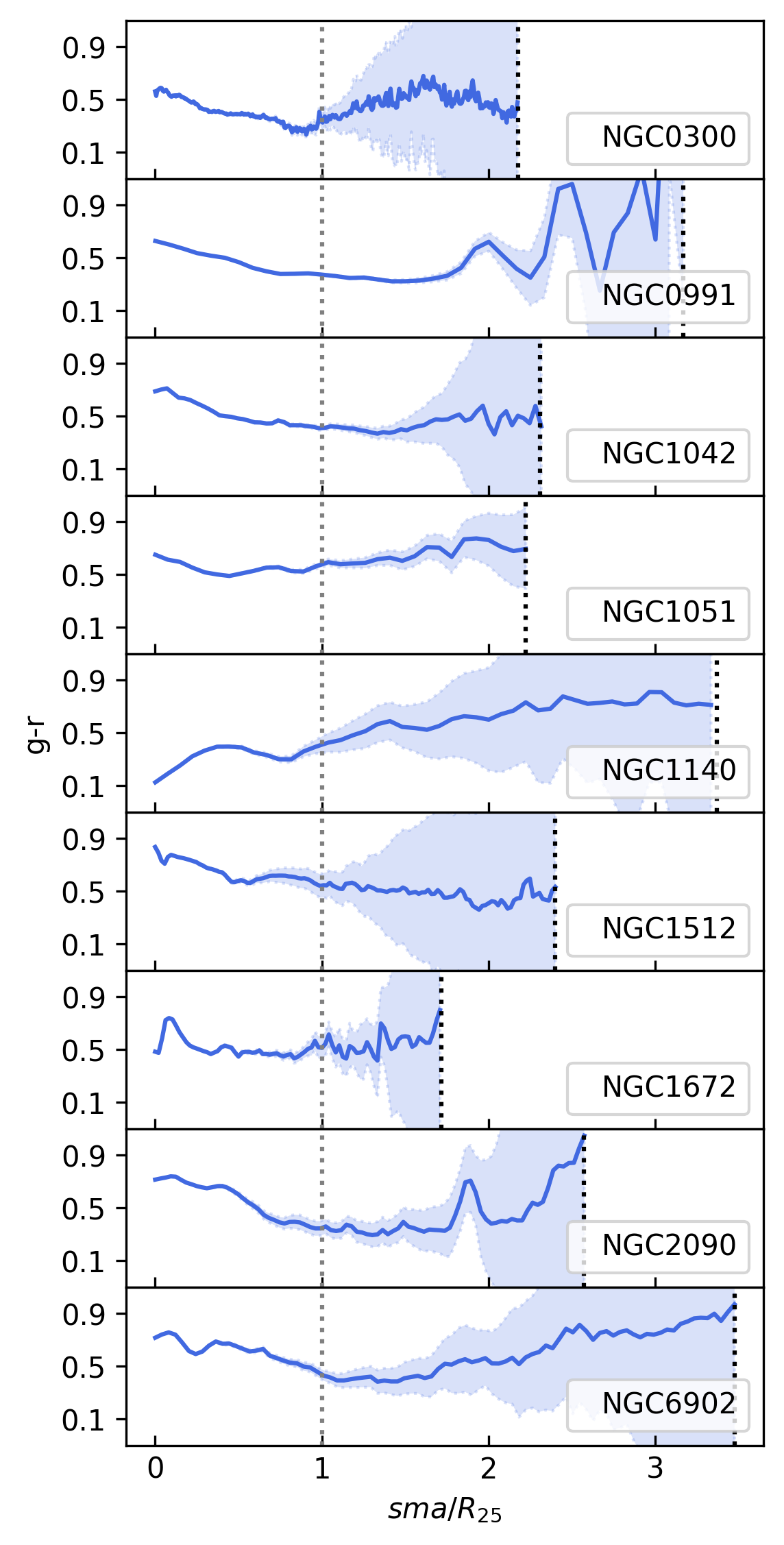}
  \caption{$g-r$ color profile of the sample. The profiles are corrected for the inclination and galactic extinction. The gray dashed line is the $R_{25}$ and the black one is the $R_{UV}$.} 
  \label{Figscolorprofile}
\end{figure}

In this section, we explore the insights that the data can offer into the nature of stellar populations present in the extended UV disks. First, we show the ($g-r$) color profiles in Figure \ref{Figscolorprofile}. As expected for star-forming galaxies, in the inner region, we obtain a negative gradient in the ($g-r$) profiles, except for NGC1140 (more details of this particular galaxy are given in Section \ref{NGC1140}). Between $R_{25}$ and $R_{UV}$, errors become increasingly large with radius and make it difficult to interpret the color profile. However, many of the profiles are constrained well enough at least in some part of the XUV region to suggest a flat blue profile (NGC991, NGC1042, NGC 2090, NGC6902), or are consistent with such a behavior (NGC300, NGC1512) within large uncertainties, with the only exception NGC1140. This suggests that in most cases the XUV region is relatively recent (and thus blue). 
In some cases, a few red peaks are found within the UV extended area (e.g., NGC0991), for which we cannot exclude some contamination missed despite careful masking. We also observe in a few cases, a reddening toward the outer edge (e.g., NGC6902, NGC1672), which could indicate that we are reaching the stellar halo population, but the uncertainties are always large enough to agree with a flat profile in those cases.

Instead of using only a color profile, we tried to use all available bands of DES and GALEX to constrain the mean age of the underlying stellar population with \textit{le Phare} \citep{1999MNRAS.310..540A,2006A&A...457..841I} using the GASPIC service\footnote{\url{https://gaspic.osupytheas.fr/}}. All details about the input parameters are given in the Appendix \ref{appSED}.

For each radius of the profiles, the best model of SFR history ($SFR_{model}$) fitting the data is given with the associated age ($Age_{fit}$) of this model. We then computed the mean stellar age ($<{age}>$) as : 

\begin{equation}\label{equationagemean} 
    <{age}> \ = \ \frac{\int_{0}^{Age_{fit}} (Age_{fit} -t) \ SFR_{model}(t) \,dt}{\int_{0}^{Age_{fit}}  SFR_{model}(t) \,dt} \ \ \ \ \  \text{in Gyr}
\end{equation}

While the code can compute uncertainties for some parameters in the Bayesian analysis, we examine the age of the stellar population, which requires knowledge of the star formation history (see Equation \ref{equationagemean}). This is given in the best fit model that remains unknown in the Bayesian analysis. Therefore, to estimate the errors, we created 100 mocks for each radius using a normal distribution of the observed uncertainties. For each of the mock, \textit{le Phare} give again a best model and age and a mean age by using the equation \ref{equationagemean}. We computed the quartiles, the median, the minimum and maximum age. In addition, we created histograms of the age distribution each 20 arcsec, (with a bin around 1 Gyr) in order to measure the mode of the distribution. We found that the distribution is wide (not bringing a strong constraints on <age>), very asymmetric, with a mode close to the value we found, but lower than the median of the mocks (see Fig. \ref{Figmean_age} of Appendix \ref{appSED}.) 
This probably arises because the models are largely degenerated with only seven bands and relatively large uncertainties in the outer parts. Nevertheless, we can try to derive broad trends, keeping these caveats in mind. 

Most of the galaxies in the sample present an age gradient in the inner region as shown in Fig. \ref{Figmean_age}, despite some variability, and are in most cases consistent with young ages in the outer region (e.g., NGC0300, NGC1042, NGC1672, NGC2090). This indication of recent formation in the outer regions is consistent with the conclusion from the color profiles (Fig. \ref{Figscolorprofile}). For several galaxies, there are peaks of older age in the outer part of the galaxies with the data or in some mock values (e.g., NGC0991, NGC1051, NGC6902). Although it is difficult to assess the precise reason, it illustrates again that some degeneracy remains with the current data. 

On the contrary, NGC1140 presents a very old stellar population in the outer disk just beyond $R_{25}$ and a very young one in the inner disc. This is consistent with the color gradient discussed above. Even if some other galaxies do present some peaks of old stellar population in some part of the extended disk (that are hard to interpret considering the uncertainties), the case of NGC1140 remains unique and is discussed in Section \ref{NGC1140}. 

Concerning the SED-fiting, le Phare provides the Chi$^2$ that are quite low, indicating that our uncertainties may be overestimated (we indeed decided to be conservative, see Sect. \ref{method}). In order to test the reproducibility of our results, we also calculated $<age>$ with Cigale \citep{2019A&A...622A.103B}, applying a similar method (and the input parameter given in appendix \ref{appSED}). The values obtained with Cigale are very similar to le Phare in most cases, with some differences sometimes at the largest radii (within the range of models found in our mocks). These results based on SED fitting must be interpreted with caution, especially due to the large uncertainties, especially in the outer regions. To improve our results in the future, we would need more bands and better images quality (especially better sky background that would allow us to reduce the uncertainties of the photometry).

\section{Discussions}\label{discussion}

\subsection{The different formation of the XUV extended structure in NGC1140}\label{NGC1140}

\begin{figure*}[t]    
    \centering
    \begin{tabular}{c c}
           \includegraphics[width=0.44\textwidth]{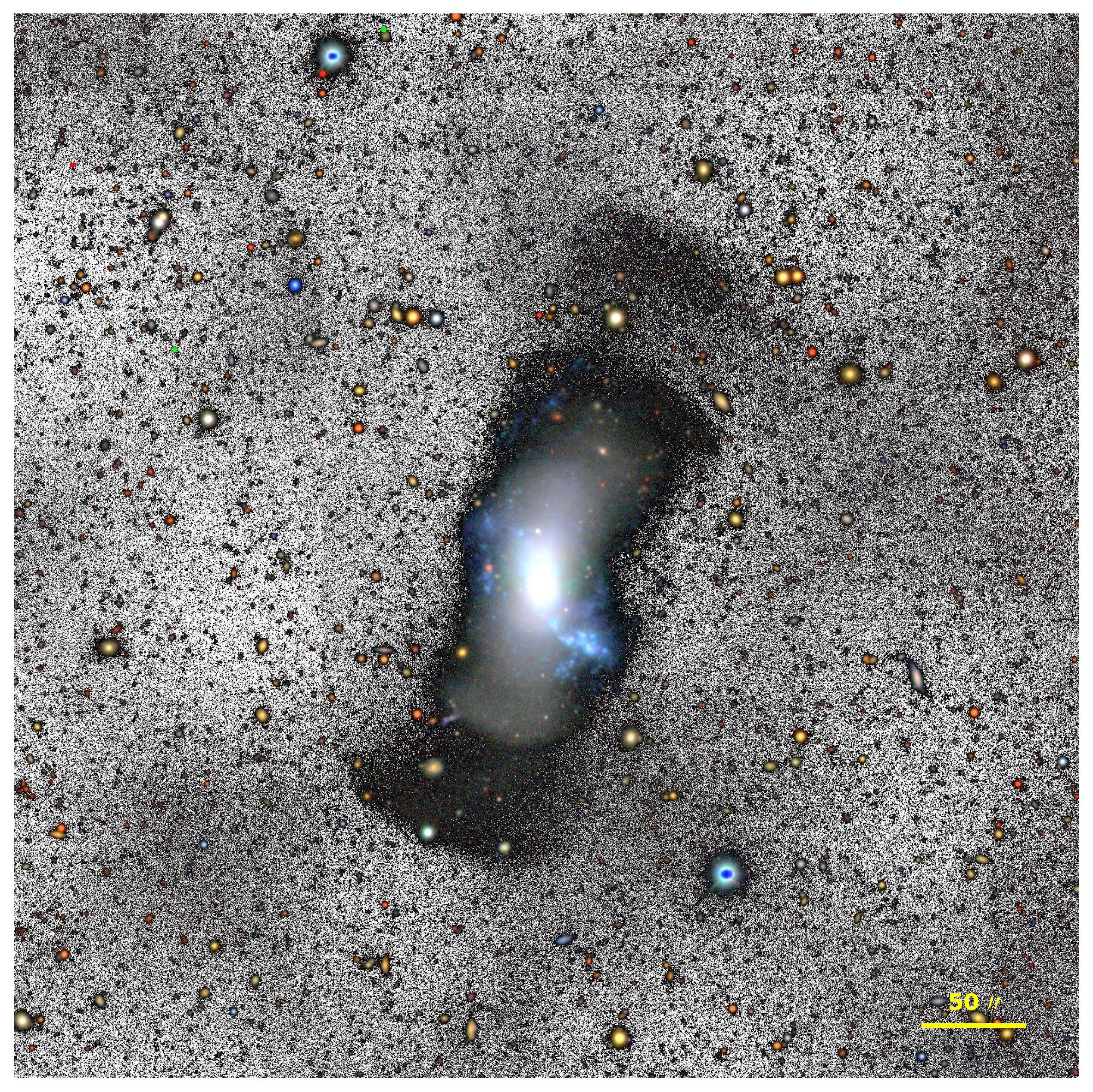}  & 
          \includegraphics[width=0.45\textwidth, height = 0.432\textwidth]{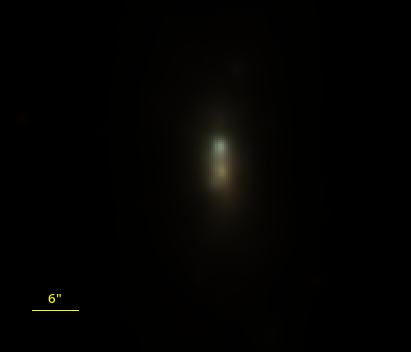} \\ 
    \end{tabular}

    \caption{NGC1140 with DES data. \textit{left} : NGC 1140 color image with $g$ band, $r$ band and $i$ band of DES. \textit{right} : zoom on the double nucleus.  }
    \label{Figure1140}  
\end{figure*}

In Sect. \ref{XUVrelation}, we could not distinguish NGC1140 from the rest of the sample based on the pure structural properties of the optical extended disc. We found it to be \emph{Malin-1-like}, with a strong anti-truncation in the XUV region. However, the color profile was found to be peculiar with a reddening from the center outward, contrary to the rest of the sample. In Sect. \ref{sedcolor}, the SED-fitting indicated the presence of an old stellar population beyond $R_{25}$, and a very recent one in the inner part, contrary to the rest of the sample again.
Inspection of the morphology of the galaxy may help to understand these differences. Indeed, the other galaxies present spiral arms and knots in the outer disc, strongly suggesting star formation in an extended gas disc, while NGC1140 exhibits a very irregular morphology, what is expected from its 
type T=9 \citep{Thilker_2007}, but also shells that are clearly visible in the DES images (see Fig. \ref{Figure1140}). We verified that we do not see more structures than the one visible in this figure, once the image is masked and binned.

Such shells usually indicate a recent merger \citep[e.g.,][and references within]{2020ApJ...891L..23Z}, even if they are more usually found in early-type systems.
Signs of mergers for NGC1140 had been already found, such as the presence and morphology of H$\alpha$ regions, and super star clusters \citep{Hunter,westmoquette10}
but the shells were not known up to now to our knowledge.
The morphology of the system is very similar to the one of VCC 848 observed by \citet{2020ApJ...900..152Z}, that was interpreted as the sign of a recent (rarely observed) dwarf dwarf merger.
Both galaxies are irregular blue compact dwarf galaxies (BCD) as established by \citet{2020ApJ...891L..23Z} and \citet{Wu2008}. In NGC1140, a double nucleus is still present (see Figure \ref{Figure1140}), while in VCC 848 there is no sign of the secondary component that has been completely disrupted \citep{2020ApJ...891L..23Z}.
It has been suggested that Ark18, with a blue extended LSB disk presenting some similitudes with GLSBs could also results from a dwarf-dwarf merger \citep{2021MNRAS.504.6179E}.

These observations of NGC1140 and VCC848 are consistent with the simulations of \citet{bekki2008} showing that the merger of dwarf rich dIrr galaxies can result in a BCD, and eventually a dIrr with spherical structures in the outer parts, which could correspond to the old ages found outward of $R_{25}$ in our case.
While studies of the mergers of dwarf galaxies are rare, NGC1140 and VCC848 are probably two very good examples, with deep optical data allowing the detection of the shells and old stellar population in the outskirts, and an extended gas distribution.
In another study on the formation of BCD galaxies via dwarf mergers, \citet{chhatkuli2022formingbluecompactdwarf} targeted a few BCDs from a catalog of dwarf merger galaxies \citep{Paudel_2018} 
on the basis of interaction features such as the tidal structure or shells seen in SDSS images. 
NGC1140 was not in this sample, as it is not in the SDSS footpath. 
\begin{table*}[t]
    \centering
    \begin{tabular}{l r r r r r }
    \hline
    Galaxy & log $M_{*}$/$M_{\odot}$ & log M$_{gas}$/M$_{\odot}$ & log SFR/(M$_{\odot}$ /yr) & log($R_{s,inner}$)/kpc.$h^-{1}$ & $\mu_{0,B}(inner)$ \\
    \hline
    NGC1140 & 8.98 &  9.34 &  0.63 & -0.23 & 20.04 \\ 
    VCC848  & 8.32 &  8.62 & -1.60 & -0.32 & 20.52 \\  
    D004    & 9.09 &  8.43 & -1.08 & -0.44 & 19.67 \\
    D036    & 8.56 &  8.87 &  ---  & -0.38 & 20.68 \\
    D047    & 8.26 &  ---  & -1.46 & -0.8  & 19.16 \\
    D055    & 9.18 &  8.19 & -0.85 & -0.15 & 20.21 \\
    D075    & 9.03 &  9.94 & -0.68 & -0.45 & 19.93 \\
    D076   & 7.92  &  8.40 &  ---  & -0.007 & 20.549 \\
    \hline
    \end{tabular}
    \hspace{1cm}
    \caption{Properties of NGC1140 and other BCD galaxies resulting from dwarf-dwarf mergers from \citet{2020ApJ...891L..23Z} and \citet{chhatkuli2022formingbluecompactdwarf}. For the stellar mass, SFR, and gas mass, we report values from the literature except for NGC1140 for which we use the results of the SED-fitting (le Phare, section \ref{sedcolor}). For $R_{s,inner}$ and $\mu_{0,B}(inner)$, we fitted the profiles provided by the authors to compare to ours. The log SFR has been converted from \citet{Kennicutt} to \citet{kroupa2001} using the conversion of \citet{refId0}.}
    \label{tabBCD}
\end{table*}

We provide in Table \ref{tabBCD} the values for the stellar mass, gas mass, SFR, and light distribution of NGC1140, VCC848 and the six BCDs of \citet{chhatkuli2022formingbluecompactdwarf}. For all galaxies in this table we computed the scale length and the central surface brightness in the inner part as in Section \ref{sbprofiles} in order to find connections between those galaxies and NGC1140. The values for NGC1140 are within the observed range for the other galaxies. The most similar galaxy is D075: both are relatively massive dwarfs, with a large gas reservoir.

\citet{2020ApJ...900..152Z} found VCC848 to be above the main sequence using \citet{shin20} as a reference. 
Our determined SFR (4.2 $M_{\odot}$/yr)  places NGC1140 even more above the main sequence, probably resulting from a temporary enhancement of star formation during the merger phase (it may be a reason for the XUV classification). \citet{westmoquette10} provided a SFR of 0.7 $M_{\odot}$/year, smaller than ours but determined from H$\alpha$, more sensitive to small timescale variations and dominated by few regions of intense star formation, while our SED-fitting is constrained by the UV emission in an extended area, sensitive to longer time-scales. For its stellar mass, NGC1140 is also one of the most gas-rich dwarfs compared to other dwarfs using \citet{galaxies10010011} as a reference, making it an extreme case in its category.

\subsection{Euclid perspectives}

Euclid offers new possibilities with a very large sky coverage and exquisite resolution. Especially the ERO (Early Release Observations) images \citep{cuillandre2024euclidearlyreleaseobservations} were processed in such away to keep the LSB signal. Among the six galaxies in the sample studied with Euclid by \citet{hunt2024euclidearlyreleaseobservations}, one (NGC2403) was classified as an XUV galaxy in \citet{Thilker_2007}. We thus performed a similar analysis to the one made for our sample.  We fit the profiles of \citet{hunt2024euclidearlyreleaseobservations} in the wide $I_e$ Euclid band, and used the GALEX images to find the $R_{UV}$ as in Sec \ref{RUV} ($R_{UV}$ = 912 arcsec and $R_{25}$ = 594 arcsec). 
To compare this galaxy observed in $\text{I}_E$ to our sample in the g band, we compared the $I_e$ profile with the profile in \citet{2012MNRAS.419.1489B} in the V band and then applied the (V-B) correction of \citet{1995PASP..107..945F}. This allowed us to add this galaxy in our figures (see Fig \ref{Figsprayberry} and Fig \ref{Figsmu0rs}).

Following the procedure of Sect. \ref{XUVrelation}, we found that NGC2403 is part of the \emph{regular} family, but with a brighter surface brightness than the three other galaxies of its category.

This shows the potential of Euclid to study XUV galaxies. \citet{urbano24} discuss the use of ERO data (and Euclid in the future) to study other LSB structures (especially traces of the past assembly of galaxies such as tails and shells). With straylight problems reduced in the future with respect to ERO data, Euclid has the potential for new discoveries in LSB science, although the impact of the different pipeline between ERO and regular observations will have to be determined.

\section{Conclusions}\label{conclusion}

For a sample of nine XUV galaxies, we used available deep optical images from DES, in combination with UV images from GALEX, in order to characterize their UV emitting outer disc, in the optical wavelength, especially to compare them to the GLSB galaxies, which are usually defined as such based on optical data (here, we call "GLSBs" as galaxies with a very large diffuseness, as defined by \citet{sprayberry_1995}, that are not necessarily very massive or extremely large).

For this, we first defined a physically based UV radius $R_{UV}$.
We then simply proposed fitting the $g$ band profile in the inner (R<$R_{25}$) and outer disk($R_{25}<R<R_{UV}$) in order to analyze the structural properties of these galaxies. 
This led us to distinguish three different families of XUV galaxies: i) the \emph{Malin-1 like} family represents the galaxies for which the outer region fit parameters places them in the GLSB region in the diagnostic figure of \citet{sprayberry_1995}, similarly to the most extreme GLSB, Malin-1 (independently of the gas content and their physical size). 
ii) The \emph{Malin-1 opposite} is a family for which the central surface brightness extrapolated from the fit of the inner region is fainter than the one of the outer disc, corresponding to a steeper slope in the outer regions than in the inner regions, contrary to the Malin-1 archetype of GLSB galaxies. iii) We dubbed \emph{regular galaxies} the rest of the sample, for which the extrapolated central surface brightness of the outer disk is fainter than the one of the inner region (like GLSBs), but the outer disk scale length is too small to fall in the GLSB regions.
XUV galaxies thus are far from being all similar to GLSB. These three families can be simply distinguished by the variable DIS (see Equation \ref{equationmurs}) or the diffuseness index of the outer part (see Fig. \ref{Figsmu0rs}), going from truncated profiles to strongly anti-truncated profiles similar to Malin-1 and the GLSB.

However, our work also demonstrated that this structural property determined with only one band is not sufficient to completely describe the outer disk for the full sample of XUV galaxies. The positive $g-r$ color gradient distinguishes one of our nine galaxies (NGC1140). Analyzes of the morphology and the stellar population strongly suggests this galaxy results from a recent merger. Its XUV nature is due to young regions, but at the same time an old stellar population is found in outer shells well visible in the DES images. The rest of the sample (8 out of 9 galaxies) presents a negative color gradient and are rather giant star forming disks, more similar to giant LSBs, but with still a variety of profiles types (anti-truncated to truncated).

In the future, many opportunities for studying XUV (and LSB) galaxies will arise. We showed that
EUCLID may allow us to study many XUV galaxies considering its large sky overage, if a pipeline adapted to extended faint objects is used like in the ERO observations, in which one XUV galaxy was observed (NGC2403). This galaxy was found to correspond to our "regular galaxy" family but is brighter than the other members. We clearly need to enlarge the samples of XUV galaxies studied in optical to verify that our proposed families are adequate to describe all XUV galaxies to better characterize them and refine our classification. Other sky surveys that may bring a large amount of data for this are the large multi-wavelength surveys that will reach depth down to 30 mag.arcsec$^{-2}$ like LSST \citep{2019ApJ...873..111I} or UNIONS\footnote{https://www.skysurvey.cc/}.
Another opportunity will be the use of dedicated telescopes like CASTLE \citep{lombardo/mnras/stz2068,Muslimov:17}, which will cover large fields with an excellent sky background, PFS, and low contamination. It will help us to characterize the outer regions of XUV galaxies and their stellar population for dedicated targets, especially the largest among XUV galaxies.

\begin{acknowledgements}

This work is partly based on tools and data products produced by GAZPAR operated by CeSAM-LAM and IAP. We thank Simona Lombardo for her help in early phases of this work, and the information she provided in the context of the CASTLE project. E.B would like to acknowledges the meaningful discussions with Mathias Urbano and his support. This work was supported by the « action thématique » Cosmology-Galaxies (ATCG) of the CNRS/INSU PN Astro. This work was partially supported by the “PHC POLONIUM” program (project number: 49136QB), funded by the French Ministry for Europe and Foreign Affairs, the French Ministry for Higher Education and Research and the Polish National Agency for Academic Exchange "POLONIUM" grant BPN/BFR/2022/1/00005.

\end{acknowledgements}
\bibliographystyle{aa} 
\bibliography{biblio}

\begin{appendix}\label{appendix}
\section{SED-Fitting}\label{appSED}

\begin{figure*}[t]
\centering          
\includegraphics[width=1\textwidth,height=1\textwidth]{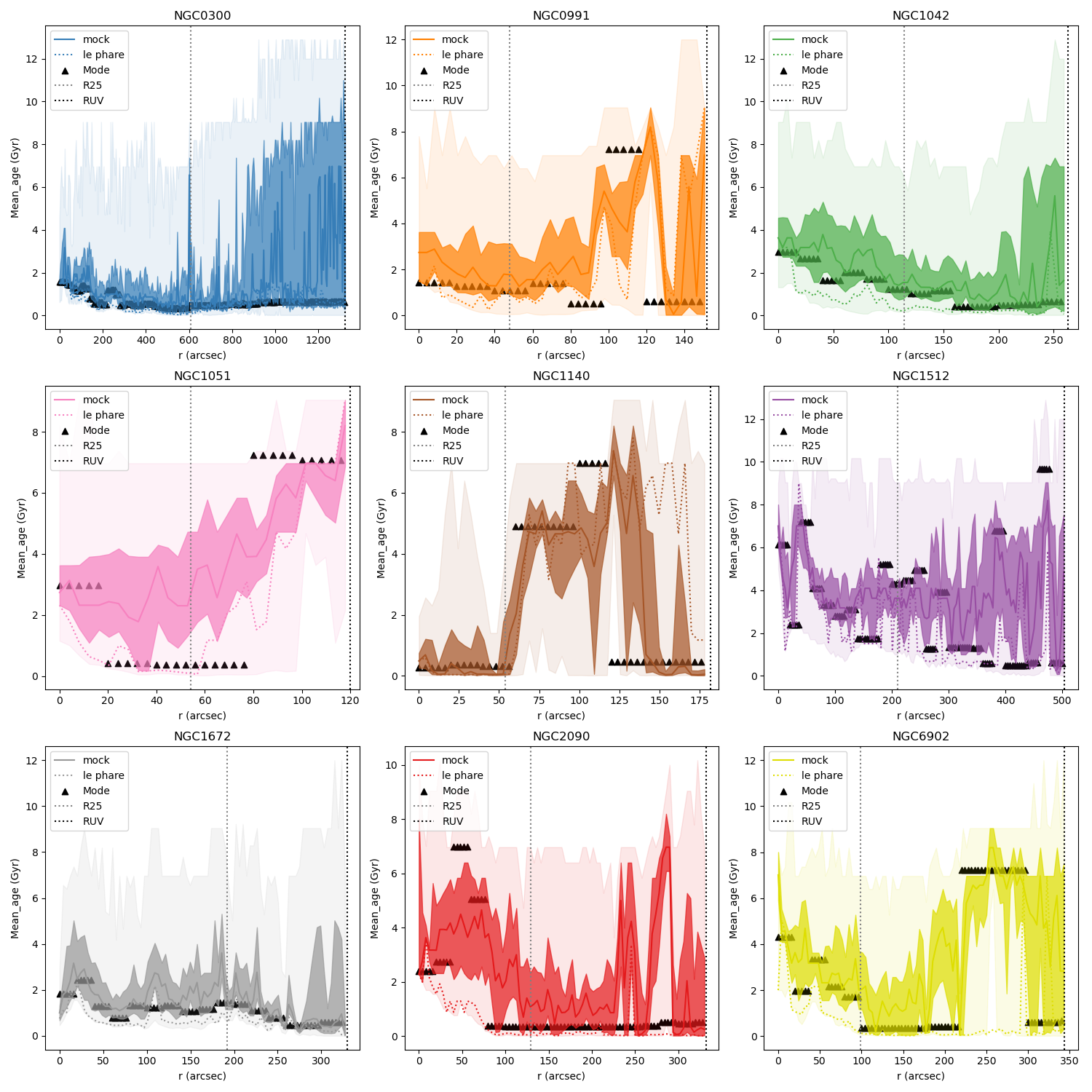}
\caption{Mean age as a function of the radius. The dotted line is the mean age calculated with Le Phare and the continuous line is the median of the mock simulations (see section \ref{sedcolor}). The triangles represent the mode of the mean age of the mocks in a range of 20 arcsec. The gray dotted line is the $R_{25}$ while the black one is the $R_{UV}$. The shadow area is the area between the first and the third quartile of the mocks while the light area is the minimum and the maximum mean age found in the mocks.}
\label{Figmean_age}
\end{figure*}

To derive the physical properties of our sources, we performed SED fitting using two tools in the astrophysical community: \textit{Le Phare} and \textit{CIGALE}. Both were accessed through the GAZPAR interface. We used the profiles described in Sec \ref{method}. It is apparent that the resolution of GALEX (about 4.5 arcsec) differs from that of DES (0.263 arcsec). However, the application of a 13x13 binning to the reprojection of GALEX onto the DES pixel results in a final pixel of 3.4 arcsec, which is comparable to the resolution of GALEX. Since we are interested in profiles along big ellipses in these large pixels, the resolution mismatch has little effect, especially at large radii where the profiles are relatively flat. We performed nevertheless a convolution of all the images with a Gaussian Kernel to match the GALEX resolution at all wavelengths. As expected, we found that the profiles remain similar but that this method introduces complexities, as some sources in the DES data that pollutes a larger area. This makes the masking process and sky determination more difficult, increasing the uncertainties in the outer part of the profiles. We thus decided to use the profiles obtained with a 3.4 arcsec binning without this convolution.

\subsection{\textit{Le Phare} Configuration}

The first modeling run was conducted using \textit{Le Phare}, as described in \citet{1999MNRAS.310..540A} and \citet{2006A&A...457..841I}. The SED fitting relies on the \citet{2003MNRAS.344.1000B} stellar population synthesis models, assuming a \citet{2003PASP..115..763C} IMF. The star formation histories are modeled with exponentially delayed model. The dust attenuation is parameterized by the redenning $E(B-V)$, varied over a grid: 0.0, 0.05, 0.1, 0.15, 0.2, 0.25, and 0.3.

We adopt a $\Lambda$CDM cosmology with $H_0 = 70\, \mathrm{km\,s^{-1}\,Mpc^{-1}}$, $\Omega_m = 0.3$, and $\Omega_\Lambda = 0.7$. The redshift range is limited to a maximum of 0.006, sampled in steps of 0.001. The DES \textit{g}-band is selected as the reference band for normalization. Default values were retained for photometric error adjustments and rescaling factors across the photometric bands. The result of this fit is shown in the Figure \ref{Figmean_age}

\subsection{\textit{CIGALE} Configuration}

We also used \textit{CIGALE} \citep{2019A&A...622A.103B} to perform an other SED fitting. This run uses the BC03 stellar population models based on \citet{2003PASP..115..763C} IMF, and a delayed star formation history module. Dust emission is treated with the \textit{dl2014} module \citep{Draine_2007}. No AGN or radio emission components are included in this analysis.

The tau main ($\tau$) for the SFH is explored in three values: 100, 5000, and 10000 Myr. We fixed only three values to constraint the age. The stellar population ages range from very young to old populations: starting with fine steps (10, 20, 50 Myr), then gradually increasing every 100 Myr between 100 and 1000 Myr, every 200 Myr between 1000 and 2000 Myr, every 500 Myr up to 5000 Myr, and finally every 1000 Myr until 13000 Myr. This allows coverage of a wide temporal baseline suitable for tracing the evolution of various stellar populations.

The color excess for young stellar populations, $E(B-V)_{\text{young}}$, spans the same grid as in \textit{Le Phare}. For the older populations, the reduction factor is set to 0.5. We adopted a low metallicity value of $Z = 0.004$, consistent with expectations for low densities in the outer regions.

Attenuation parameters include a fixed UV bump amplitude of 3.0 and an attenuation law slope of 0. All other parameters were left unchanged from their default settings in \textit{CIGALE}.

\end{appendix}

%
%

\end{document}